\documentclass[a4paper]{article}
\usepackage{INTERSPEECH2022}
\usepackage{amsmath,graphicx}
\input{def.set}
\usepackage{makecell}
\usepackage{xcolor,colortbl}
\usepackage{color}
\usepackage{colortbl}

\usepackage{multirow,bigdelim}
\usepackage{bm}
\usepackage{subfigure}
\usepackage{nccmath}
\usepackage{float}

\usepackage{amsmath}
\usepackage{hyperref}
\usepackage{caption} 
\usepackage{booktabs}
\usepackage{bbding}
\usepackage{pifont}
\usepackage{wasysym}
\usepackage{amssymb}
\usepackage{graphicx}
\usepackage{booktabs,amsfonts,dcolumn}

\usepackage{algorithmic, algorithm}
\usepackage{mathtools}
\usepackage{tipa}

\DeclarePairedDelimiter{\ceil}{\lceil}{\rceil}

\newcommand{\zhenk}[1]{{\color{black}{#1}}}
\newcommand{\hieng}[1]{{\color{black}{#1}}}

\usepackage{scalerel,stackengine}
\usepackage{algorithmic, algorithm}

\usepackage{caption} 
\usepackage{booktabs}
\usepackage{bbding}
\usepackage{pifont}
\usepackage{wasysym}
\usepackage{amssymb}
\usepackage{graphicx}
\usepackage{booktabs,amsfonts,dcolumn}

\usepackage{algorithmic, algorithm}
\usepackage{mathtools}
\usepackage{tipa}

\title{Sub-8-Bit \hieng{Quantization Aware Training} for 8-Bit Neural Network Accelerator \\\hieng{with} On-Device \hieng{Speech Recognition}}

\name{\begin{tabular}{@{}c@{}}
		Kai Zhen$^{ \dagger}$ \qquad 
		Hieu Duy Nguyen$^{\dagger}$ \qquad 
		Raviteja Chinta$^{\star}$ \qquad \\ 
		Nathan Susanj$^{ \dagger}$ \qquad 
		Athanasios Mouchtaris$^{ \dagger}$ \qquad 
		 Tariq Afzal$^{\star}$ \qquad 
		Ariya Rastrow$^{\dagger}$
\end{tabular}}

\address{Alexa AI, Amazon, USA$^{\dagger}$\\
Hardware Compute Group, Amazon, USA$^{\star}$\\
~\texttt{\{zhenk, hieng, ravitec\}@amazon.com}\\
~\texttt{\{nsusanj, mouchta, tafzal, arastrow\}@amazon.com}
}

\begin{document}
%
\maketitle
\begin{abstract}
We present a novel sub-8-bit \zhenk{quantization-aware training (S8BQAT) scheme} for 8-bit neural network accelerators.
Our method is inspired from Lloyd-Max compression theory with practical adaptations for a feasible computational overhead during training. With the quantization centroids derived from a \zhenk{32-bit} baseline,  we augment training loss with a Multi-Regional Absolute Cosine (MRACos) regularizer that aggregates weights towards their nearest centroid, effectively acting as a pseudo compressor. Additionally, a periodically \hieng{invoked} hard compressor is introduced to improve the convergence rate by emulating \hieng{runtime} model weight quantization. 
\hieng{We apply S8BQAT on speech recognition tasks using Recurrent Neural Network-Transducer (RNN-T) architecture. 
With S8BQAT, we are able to increase the model parameter size to reduce the word error rate by 4-16\% relatively, while still improving latency by 5\%.}


\end{abstract}
%

\section{Introduction}
\label{sec:intro}

\zhenk{Latency reduction without introducing noticeable accuracy degradation is critical to on-device automatic speech recognition (ASR) systems in various scenarios, such as in in-car units and on portable devices, where  Internet connectivity can be intermittent.}
Ubiquitous as end-to-end ASR models are \cite{song2021non,gulati2020conformer, chang2021end, radfar2020end}, deploying them on edge can be challenging due to the strict limit of bandwidth and memory of edge devices \cite{sainath2020streaming, kim21m_interspeech}. Hence, it is highly necessary to improve model efficiency, i.e. model memory size vs. model accuracy, through applying model optimization and compression techniques when running real-time ASR on-device.




Efficient neural network inference can be approached by various ways. The teacher-student training paradigm for knowledge distillation simplifies the network topology, although the student may fail to mimic the teacher's behavior if the model capacity is too low \cite{cho2019efficacy}. Furthermore, it usually requires additional network compression techniques prior to the hardware deployment \cite{gholami2021survey}. Alternatively, enforcing model sparsity can significantly reduce memory footprint \cite{zhu2017prune, zhen2021sparsification}. However, one can only realize the inference speedup if the sparsity pattern matches the specific memory design of the hardware \cite{stamenovic2021weight}.

\hieng{Neural network quantization can be effectively employed to compress 32-bit weights down to 8-bit, or activations to 5-bit \cite{zhen19_interspeech, YangH2021sanac}, via applying a simple post-training quantization step \cite{wang2020towards, shomron2021post} or a more involved QAT mechanism \cite{bhandare2019efficient, nguyen2020quantization}.
}
Several quantization methods have been proposed to lower the bit-depth to 4 \hieng{\cite{zafrir2019q8bert, lou2019autoq, banner2018post}}, or even 1 for proof-of-concept bit-wise neural networks \cite{kim2019incremental}.
Nevertheless, these methods can be heuristic which rely on an extensive hyper-parameter tuning to maintain the accuracy level. 
\zhenk{Furthermore, such sub-8-bit quantization methods require sub-8-bit operators on neural network accelerators (NNAs)}, which often have inferior performance compared to their 8-bit counterpart due to the reduced numerical accuracy. Consequently, sub-8-bit NNAs are less adopted and thus there is no real latency measurement for existing sub-8-bit approaches \cite{bhandare2019efficient, zafrir2019q8bert, banner2018post, kim2019incremental}. The most prevalent type of NNAs are rather based on 8-bit arithmetic operators, i.e. addition/multiplication/etc, accepting 8-bit inputs and computing the outputs via bitwise operations. For the rest of the paper, we will refer to this type as INT8 NNA.    

In this work, we propose a novel sub-8-bit quantization \hieng{aware training (S8BQAT)} that integrates with INT8-based runtime NNAs. \zhenk{It differs from our previous linear-QAT method \cite{nguyen2020quantization} in twofold.}
Firstly, S8BQAT distills quantization centroids from  \zhenk{a pre-trained 32-bit} baseline via \hieng{a mechanism derived from} Lloyd-Max scalar quantization theory. 
We introduce Multi-Regional Absolute Cosine (MRACos) regularizer, which is INT8 compatible and computationally efficient.  The MRACos regularizer penalizes off-the-centroid \hieng{weights and aggregates them} towards their nearest quantization centroids. Additionally, the MRACos regularizer is accompanied by a periodic compressor that assigns each model weight to that nearest quantization centroid, ensuring quantization convergence and therefore \hieng{minimizing runtime} quantization-induced performance degradation.
\hieng{Throughout the paper, we also refer to the MRACos regularizer and periodic compressors as soft and hard compressors, respectively. 
The soft compressor affects the gradient calculation only through the use of the regularization term, while the hard compressor quantizes each model weight to the exact centroid that are also used at runtime inference.
We apply our S8BQAT into the ASR task and measure the performance} in terms of word error rate (WER) and \hieng{on-device runtime user} perceived latency (UPL) on various runtime settings. 
\hieng{Results show that the proposed S8BQAT achieves superior WER-UPL tradeoff compared to an 8-bit baseline. In particular, we increase the number of model parameters by 10.3\%, thus reducing WER by 4-16\% relative while reducing UPL by 5\% with S8BQAT.}

\hieng{The rest of the paper is structured as follows. The proposed S8BQAT and the associated data loading mechanism are introduced in} Sec.\ref{sec:algo}. In Sec.\ref{sec:exp}, we conduct runtime validation including the measurement of model accuracy and on-device UPL from general-purposed RNN-T models for speech recognition, along with an ablation study. Sec. \ref{sec:con} \hieng{concludes the paper}.

\section{Algorithm description}
\label{sec:algo}
\subsection{Sub-8-bit data loading mechanism in NNA}
The proposed algorithm \hieng{compresses} deep learning models that are hosted on an NNA.
At runtime, the NNA loads the model weights from the system memory into the neural computing unit's local memory buffer (data moving phase) to perform bitwise arithmetic operations. This data moving phase needs to be accounted for each model inference call as often NNAs have limited on-chip memory to fully cache model weights locally.

 \begin{figure}[t]
	\centering
	{\includegraphics[width=\columnwidth]{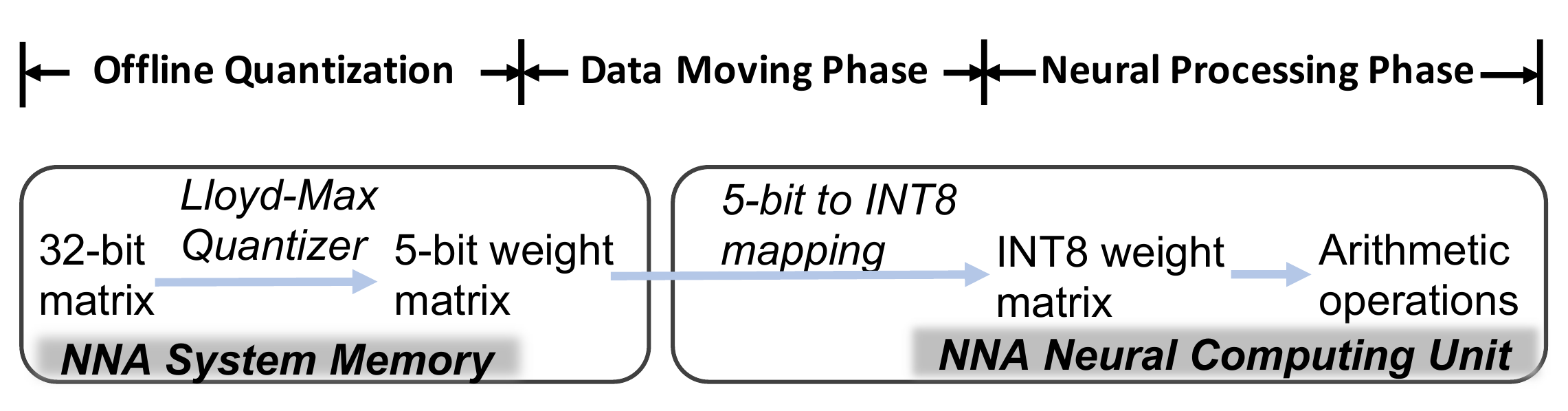}}
	\vspace{-0.17in}
	\caption{NNA loads 32-bit model weights in the sub-8-bit form and decompresses them for integer arithmetic operations.}
	\label{fig:nna}
	\vspace{-0.17in}
\end{figure}
High efficiency can be achieved by accelerating matrix related operations, such as matrix multiplication on NNAs. Yet, the data moving phase is time consuming 
due to the constrained available memory bandwidth on-device.
\hieng{To reduce the bandwidth and consequently the latency, we quantize} model weights into sub-8-bit (offline quantization), then transfer them to the NNA's neural computing unit where sub-8-bit weights are decompressed into INT8 format for the neural processing phase (see Figure \ref{fig:nna}). \hieng{Consider a weight matrix with the shape of $(1024, 4096)$. With Lloyd-Max quantizer representing all weights by 32 distinct values or 5-bit, the compressed matrix requires 2.5MB, instead of 4MB for 8-bit, thus reducing the in-memory size and data transfer latency by 37.5\%}.



\subsection{Lloyd-Max scalar quantization \hieng{theory}}
\hieng{The problem of compressing a set of weights into another set with a smaller cardinality is solved by S. Lloyd and J. Max, which is often referred to as Lloyd-Max scalar quantization theorem \cite{lloyd1982least, max1960quantizing}. }
Let $\{c_1, ..., c_k\}$ be $k$ partitions of the model weights $\bw = \{w_1, ..., w_n\}$, and $\{m_1, ..., m_k\}$ be the prototypes (or quantization centroids) for the corresponding $k$ partitions.
Lloyd Max algorithm minimizes the mean squared error between the model weights and corresponding centroids:
$\mathcal{H} = \sum\limits_{k=1}^K\sum\limits_{w_i\in c_k}||w_i-m_k||^2$. \hieng{The solutions of $\{m_1, ..., m_k\}$ can be derived in closed-form \cite{lloyd1982least, max1960quantizing}, or via an iterative method which is also used for K-Means clustering \cite{macqueen1967some}:} for each iteration, we $1)$ assign each $w_i$ to the nearest $m_k$: $c_k^{\text{new}} = \{w_i: \arg\min\limits_{k'}||w_i-m_{k'}||^2=k\}$; and $2)$ update prototypes by setting $m_k^{\text{new}} = \frac{1}{|c_k^{\text{new}}|}\sum\limits_{w_i\in c_k^{\text{new}}}w_i$, where both steps decrease $\mathcal{H}$ unless the algorithm has converged.
With all weights $w\in c_i$ represented by $\{m_1, ..., m_k\}$, they are quantized into $\ceil*{\log_2k}$ bits. 

%
%
%

\subsection{Lloyd-Max quantizer-like regularization}
Directly applying Lloyd-Max scalar quantizer to sub-8-bit QAT can be problematic. Firstly, the quantization centroids are not guaranteed to conform to INT8 format. Consequently, at \hieng{runtime}, when they are updated to $k/128$, where the integer $k\in[-128, 127]$, the model performance 
is subject to degradation.
Secondly, executing the Lloyd-Max algorithm per training step is computationally expensive and memory consuming. 

\begin{figure}[t]
\centering
 \label{fig:kmean-reg}
\subfigure[Multi-regional absolute cosine 
regularizer with 31 peaks]{\includegraphics[width=0.49\textwidth]{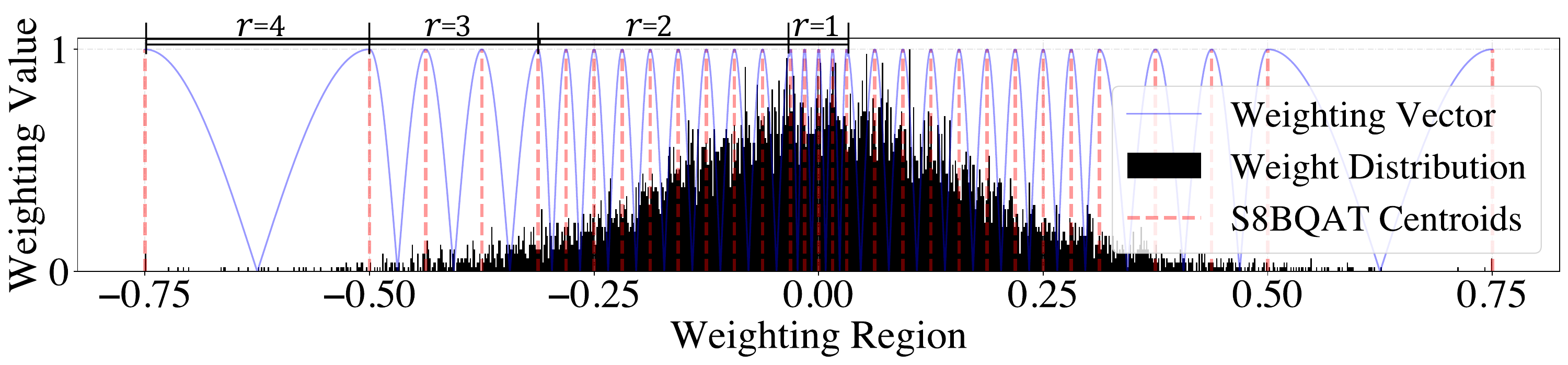}}
\vspace{-0.17in}
\subfigure[Gradient decays when the cosine frequency gets smaller]{\includegraphics[width=0.49\textwidth]{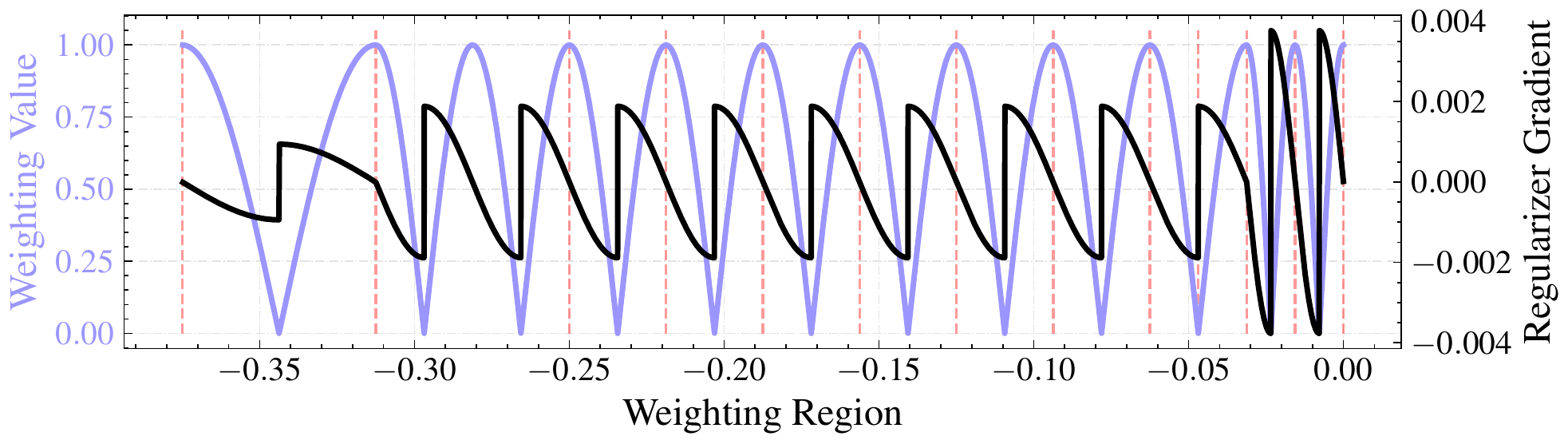}}
\hspace{0.0in}
\caption{Proposed soft compressor and its gradient}
 \label{fig:acos-reg}
 \vspace{-0.17in}
 \end{figure}

To mitigate this issue,  we propose multi-regional absolute cosine (MRACos) regularizer
\begin{align}
\mathcal{L}_\text{MRACos-reg}(\bw) = \sum_{w_i\in \bw} \sum\limits_{r=1}^R \delta_r\lambda_r(1-|cos(\pi\theta_r w_i)|),
\label{eq:mr-acos}
\end{align}
\zhenk{where $|cos(\pi\theta_r w_i)|$, $\lambda_r$ and $\theta_r$ define the weighting vector, \hieng{regularization} weight coefficient and frequency of the cosine function in region $r$,} while 
\hieng{$\delta_r(w_i)$=$1$ \zhenk{for all $w_i\in\bw$ that are} inside the range of the r-th region, and \zhenk{$\delta_r(w_i)$=$0$} otherwise.
It approximates the Lloyd-Max quantization centroids 
\zhenk{per region, e.g. $r$=$3$ in Figure \ref{fig:acos-reg} (a)}. 
Note that the frequency of the cosine function for all regions in Eq.\ref{eq:mr-acos}
is chosen such that regularizer maxima have INT8 format $k/128$, in which $k \in [-128, 127]$, and closest to the Lloyd-Max quantizer.}
The weighting vector
penalizes model weights by how far they are off the nearest centroid: as shown in Eq.\ref{eq:mr-acos}, the further the model weight is off, the larger penalty is applied; the weight receives no penalty when it's on the centroid, as it leads to no extra degradation when being quantized. 
\zhenk{Weights outside all regions are clipped.}

The gradient of MRACos regularizer can be calculated as,
\begin{align}
    \frac{\partial \mathcal{L}_\text{MRACos-reg}}{\partial w_i} =
    \sum\limits_{r=1}^R \theta_r\lambda_r (-1)^{\xi} \pi\delta_r sin(\pi\theta_r w_i),
    \label{eq:derivative}
    \vspace{-0.17in}
\end{align}
in which 
$\xi$=$0$ if $w\in[\frac{t-2}{\theta_r}, \frac{t}{\theta_r}]$ with $t$=$\{\cdots, -3, -1, 1, 3, \cdots\}$ and $\xi$=$1$, otherwise.
Note that the cosine frequency $\theta_r$ in Eq. \ref{eq:derivative} becomes a decay factor in the gradient of the regularizer. 
For a region with a relatively small cosine frequency, the gradient there will be comparably small (see Figure \ref{fig:acos-reg} (b)). 
Admittedly, with a relatively large $\lambda_r$, weights will be more aggressively aggregated toward quantization centroids for \hieng{faster quantization convergence}. However, 
\hieng{setting a high regularization weight will negatively affect the model performance, as discussed in Sec.\ref{sec:exp}.}


 \begin{figure}[t]
	\centering
	{\includegraphics[width=\columnwidth]{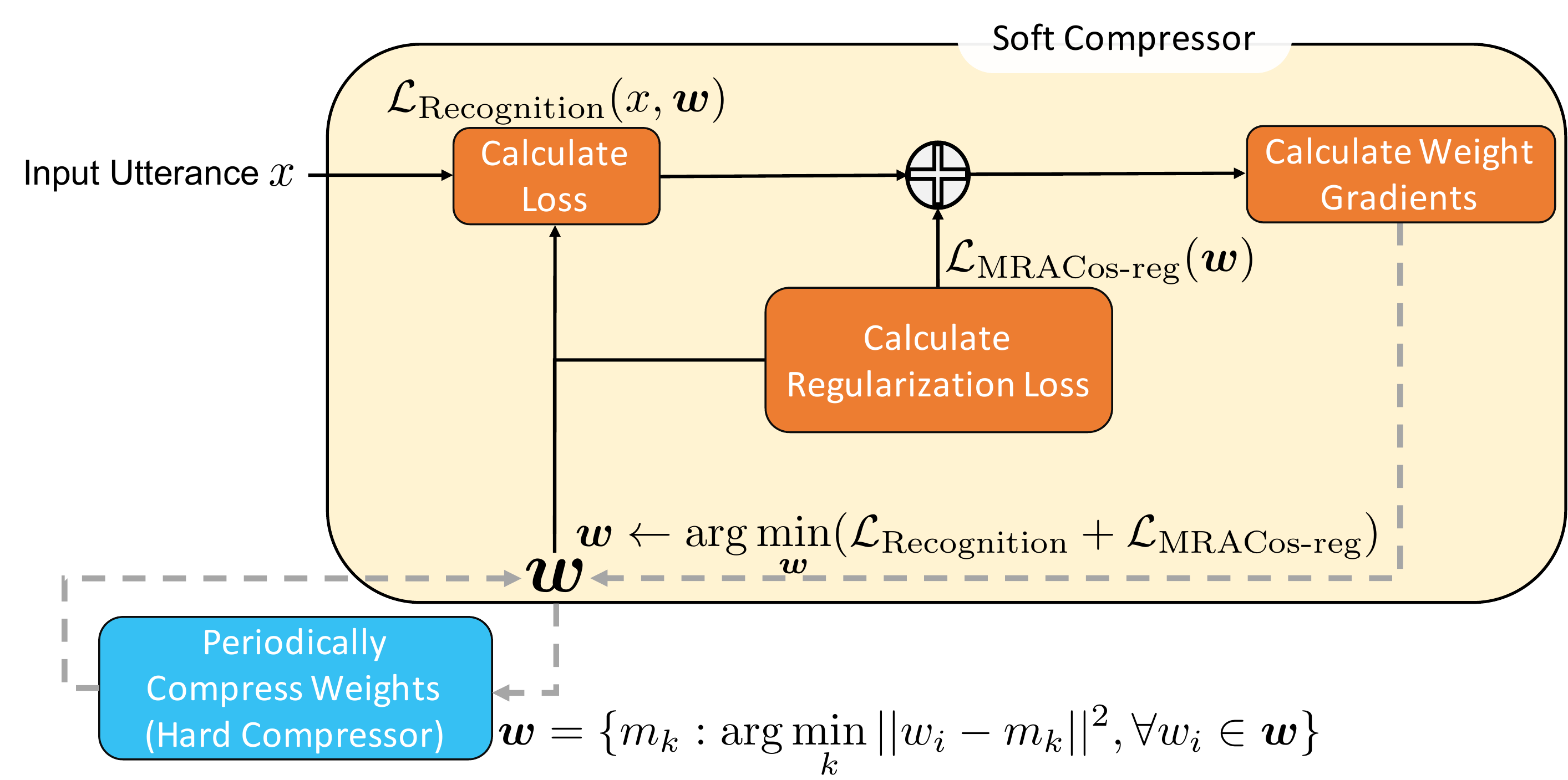}}
	\vspace{-0.17in}
	\caption{Data flow during training with \hieng{S8BQAT: the soft compressor operates at the gradient level while the hard compressor mimics runtime/post-training quantization.}}
	\label{fig:soft-and-hard}
	\vspace{-0.17in}
\end{figure}

\subsection{Periodic hard compressor}
Instead of increasing the \hieng{regularization} weight 
to address the gradient decay issue,
we introduce a periodic hard compressor that performs \hieng{runtime} quantization during model training per $\tau$ epochs (Figure \ref{fig:soft-and-hard}).
We measure the quantization convergence rate of weights for the $k$-th partition $c_k$, and its centroid $m_k$ as 
\begin{align}
   \gamma_k = \frac{|||w_i-m_k||^2<\epsilon|}{|w_i \in c_k|}, 
\end{align}
 where the threshold $\epsilon$ is relatively small. \zhenk{As discussed in Sec.\ref{sec:exp}, $\gamma_k$ can be boosted effectively with the hard compressor to reduce the quantization induced degradation at runtime.}

 


\begin{table}[!t]
	\centering
		\caption{Architecture for M-II and M-III. S8BQAT can be enabled layer-wise: layers of M-II in gray are with 5-bit S8BQAT while others in 8-bit. M-III applies 5-bit S8BQAT to all layers.}
	\scriptsize
	\setlength\tabcolsep{4.pt}
		\label{tab:topo}
\begin{tabular}{ c| c|c|c|c }
 \toprule
 & Layer &Kernel & Recurrent Kernel & \#Params\\
 \midrule
\multirow{5}{*}{\rotatebox[origin=c]{90}{Encoder}} & \begin{tabular}{@{}c@{}}L0\\\end{tabular} & (192, 4096) & (1024, 4096) & \\
& \begin{tabular}{@{}c@{}}\cellcolor{gray!25}L1\end{tabular} & \cellcolor{gray!25}(1024, 4096) & \cellcolor{gray!25}(1024, 4096) & \\
& \begin{tabular}{@{}c@{}}\cellcolor{gray!25}L2\end{tabular} & \cellcolor{gray!25}(2048, 4480) & \cellcolor{gray!25}(1120, 4480) &47.8 \\
& \begin{tabular}{@{}c@{}}\cellcolor{gray!25}L3, L4\end{tabular} & \cellcolor{gray!25}(1120, 4480) & \cellcolor{gray!25}(1120, 4480) & \\
& \begin{tabular}{@{}c@{}}\cellcolor{gray!25}Projection\end{tabular} & \cellcolor{gray!25}(1120, 512) & -- &  \\
\midrule
\multirow{3}{*}{\rotatebox[origin=c]{90}{Decoder}} & \begin{tabular}{@{}c@{}}L0\\\end{tabular} & (512, 4352) & (1088, 4352) \\
& \begin{tabular}{@{}c@{}}\cellcolor{gray!25}L1\end{tabular} & \cellcolor{gray!25}(1088, 4352) & \cellcolor{gray!25}(1088, 4352) &17.0  \\
& \begin{tabular}{@{}c@{}}\cellcolor{gray!25}Projection\end{tabular} & \cellcolor{gray!25}(1088, 512) & -- & \\
\midrule
\multirow{3}{*}{\rotatebox[origin=c]{90}{Jointer}} & \begin{tabular}{@{}c@{}}L0\\\end{tabular} & (512, 2501) & -- & \multirow{3}{*}{{2.6}}\\
& \begin{tabular}{@{}c@{}}\cellcolor{gray!25}Decoder \\\cellcolor{gray!25}Embedding\end{tabular} & \cellcolor{gray!25}(2501, 512) & -- & \\
\bottomrule
\end{tabular}
\vspace{-0.17in}
	\label{tab:rnn-t}
\end{table}

\begin{table*}[!t]
\normalsize
	\centering
	\captionsetup{width=\linewidth}
	\caption{Accuracy and latency comparison between 8-bit quantization and S8BQAT quantization: the proposed S8BQAT mechanism allows the ASR model to increase the number of parameters to improve accuracy and reduce latency, simultaneously. Here, P50 and P90 denote the 50\% and 90\% percentile latency for all utterances in the test set.}
	\setlength\tabcolsep{1.5pt}
	\resizebox{2.1\columnwidth}{!}
	{
		\begin{tabular}{  c | c c|c  c | c  c | c  c | c  c | c c|c c}
			\toprule
			\multicolumn{1}{ c | }{}         &
			\multicolumn{2}{ c | }{Model Specification}         &
			\multicolumn{8}{ c | }{Normalized ASR WERs}         &
			\multicolumn{4}{ c  }{Normalized User Perceived Latency}
			\\
			\cmidrule(r){2-15}
			\multicolumn{1}{ c | }{}         &
			\multirow{2}{*}{\#Param}&
			\multirow{2}{*}{\shortstack{Quantization \\ Mode}} &
			\multicolumn{2}{ c | }{Frequent} & \multicolumn{2}{ c | }{Entertainment} & \multicolumn{2}{ c | }{Appliances} & \multicolumn{2}{ c | }{Rare} & \multicolumn{2}{ c  }{} \\  \cmidrule(r){4-15}
			\multicolumn{1}{ c | }{}     &
			&
			&
			WER      &  Rel. Dgrd.     & WER     &   Rel. Dgrd.& WER     &   Rel. Dgrd.       &   WER   &   Rel. Dgrd.           &     P50 &   Rel. Dgrd. & P90 &   Rel. Dgrd.   \\ \midrule
			
			\multirow{1}{*}{M-I} &61.0M &8-bit
&1.00 & - & 2.09 & - & 2.24 & - & 3.24 & - & 1.00 & - & 1.48 & -\\
\multirow{1}{*}{M-II} &67.3M & \hieng{8/5-bit}     
& 0.84 & -0.16 & 1.93 & -0.08 & 2.14 & -0.04 & 3.10 & -0.04 & 0.95 & -0.05 & 1.43 & -0.03 \\
\multirow{1}{*}{M-III} &67.3M & 5-bit 
& 0.84 & -0.16 & 1.94 & -0.07 & 2.13 & -0.05 & 3.12 & -0.04 & 0.93 & -0.07 & 1.41 & -0.05 \\ 
			\bottomrule
			
		\end{tabular}
	}
	\label{tab:comp}
\end{table*}

\section{Experiment}
\label{sec:exp}
\subsection{Experimental setup}
\hieng{To validate the effectiveness of the proposed S8BQAT method, we apply it into the ASR task using the Recurrent Neural Network - Transducer (RNN-T) \cite{graves2013speech} architecture and measure two performance metrics: the word error rate (WER) for accuracy and user perceived latency (UPL)}.
 We consider several RNN-T variants, and compare the performance against a linear quantization-aware training baseline method \cite{nguyen2020quantization}. Essentially, the proposed method differs from its baseline in two aspects: the soft compressor approximates Lloyd-Max scalar quantization to regularize model weights in a non-linear space while the hard compressor is introduced and invoked periodically to emulate \hieng{runtime} quantization.
 


We consider 3 RNN-T model settings, all with 5 encoding layers, 2 decoding layers, 1 joint layer and 2.5k word pieces: M-I is the baseline including 61.0M parameters, and trained with the 8-bit linear QAT mechanism \cite{nguyen2020quantization}. M-II and M-III adopt the topology of M-I but with a slightly larger number of hidden units: 
\hieng{specifically, the number of hidden units are increased from 1024 to 1120 and 1088 for Encoder L2-L4/Projection layer and Decoder, respectively (see Table \ref{tab:rnn-t}).}
Consequently, the number of parameters for M-II and M-III is 67.3M, or 10.3\% larger than M-I. 
\hieng{
Under S8BQAT, the process is as follows. 
We first train M-II and M-III for a pre-determined duration, e.g. 10k out of 850k steps; then extract the weight distribution and sub-8-bit configurations for all layers of M-III, and all but L0 layers of M-II; then apply the proposed S8BQAT mechanism (see Figure \ref{fig:soft-and-hard}) to train M-II and M-III for the rest of the epochs; and finally, compress all sub-8-bit layers of M-II and M-III after the training finishes.}


M-I, II and III are trained with a de-identified far-field dataset including 100k hours of human transcribed utterances and 40k hours of utterances generated by self-supervised learning speech model. 
\hieng{All models are trained for 850k steps. We consider 4 far-field test sets for the WER evaluation: Frequent, Appliances, Entertainment, Rare, each with 50K utterances. Frequent and Rare datasets include tasks that are frequently or rarely queried, while Appliances and Entertainment contain queries from specific supported Appliances and Entertainment tasks.} We benchmark latency metrics on 6k test utterances.

For reproducible purposes, we also train RNN-T models with S8BQAT on LibriSpeech corpus \cite{panayotov2015librispeech} with 960k hours of training data for 120k steps, and measure WERs with 5.4k, 5.3k, 5.4k, and 5.1k hours of dev-clean, dev-other, test-clean and test-other data, respectively. 

\subsection{Experimental results}

\begin{figure*}[t]
	\centering
{\includegraphics[width=\linewidth]{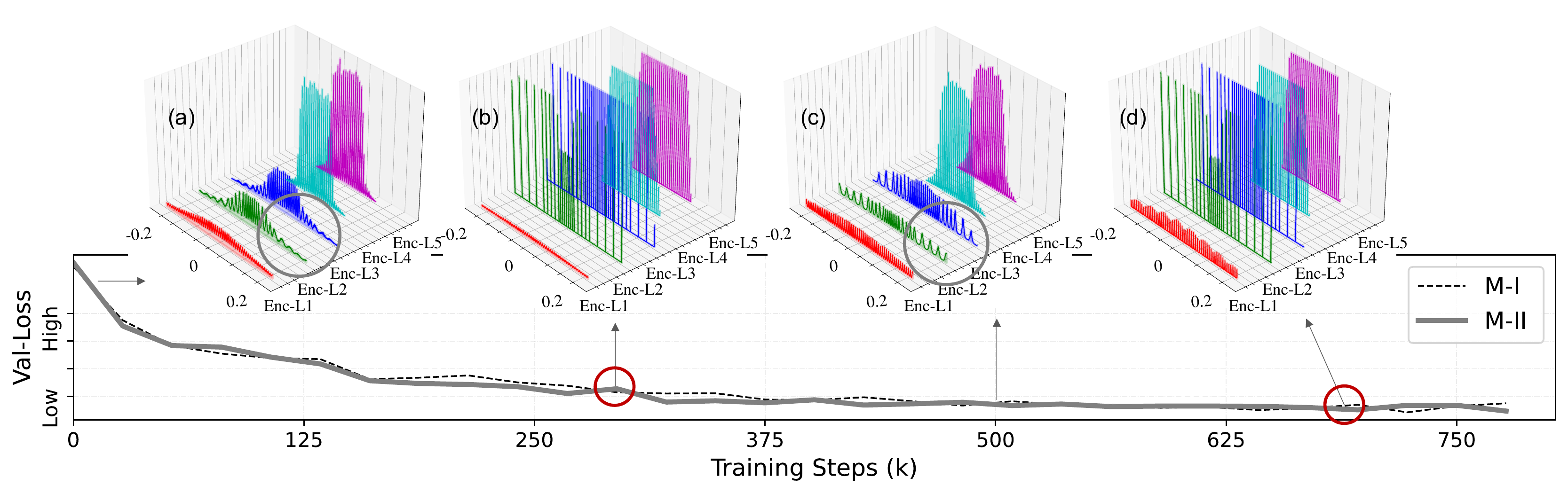}}
	\vspace{-0.15in}
	\caption{The during-training behaviour of proposed S8BQAT algorithm (M-III is not shown to save space): the weights are optimized to lower the quantization loss defined in Eq.\ref{eq:mr-acos} with the hard compressor periodically performed to maximize the quantization convergence rate. For simplicity, only the progress for the weight histogram of RNN-T encoder is shown.}
	\label{fig:big-wide}
\end{figure*}

\subsubsection{Accuracy and latency analysis}
We compare the accuracy and latency performance among M-I, M-II and M-III in Table \ref{tab:comp}. \hieng{We normalize the WER numbers by the WER of M-I for Frequent test set. The latency numbers are normalized by the latency of M-I for latency test set. The latency} is observed to be reduced even when the number of parameters increases thanks to the employment of S8BQAT: concretely, compared to the 8-bit baseline (M-I), M-II lowers the UPL-P50 by 5\% and UPL-P90 by 3\%; a more aggressive quantization configuration in M-III further brings down the latency with 6\% UPL-P50 reduction and 5\%  UPL-P90 reduction, respectively. Moreover, M-II and M-III consistently show the accuracy improvement over the 8-bit baseline from all 4 test sets, thanks to their slightly increased number of parameters. 

The mechanism of S8BQAT is depicted in Figure \ref{fig:big-wide}, where we extract 4 snapshots of M-II during training and plot their weight distributions. With the MRACos regularizer, weights are effectively driven to quantization centroids even with only 50k training steps (Figure \ref{fig:big-wide} (a)). 
However, peaks in the tail of the distribution (grey circle in Figure \ref{fig:big-wide} (a)) are far less spiky (indicating a lower quantization convergence rate, $\gamma$).
\hieng{This is mitigated at step 500k (see Figure \ref{fig:big-wide} (c)), as all 5-bit regularized weights are hard quantized via the periodic hard compressor at step 350k (see Figure \ref{fig:big-wide} (b))}. Had the hard compressor not been invoked, the convergence rate at the tail of the weight distribution would have still been low, due to the decayed gradient issue as discussed in Sec.\ref{sec:algo}. 
\hieng{On the other hand, utilizing only the hard compressor results in large performance fluctuation and fundamentally worse WERs \cite{nguyen2020quantization}. Thanks
to the quantization-aware training with MRACos regularizer, the weights are gradually pushed towards the quantization centroids, leading to smaller fluctuations in the validation loss (see Figure \ref{fig:big-wide} (d)).}

\begin{table}[!t]
	\centering
	\caption{QAT methods comparison and ablation study on LibriSpeech datasets: by default, $\lambda$=$5e$-$4$ and $\tau$=$35$ for S8BQAT.}
	\footnotesize
	\setlength\tabcolsep{4.pt}
		\begin{tabular}{ l | c | c|c|c}
			\toprule
			& dev-clean  & dev-other& test-clean & test-other
			\\
			\midrule
			32-bit baseline & 8.11 & 21.27 & 8.68 & 22.29 \\
            8-bit (linear-QAT) & 8.15 & 21.41 & 8.70 & 22.36 \\
			5-bit (linear-QAT) & 9.60&  23.05&  9.76 & 24.10 \\
			4-bit (linear-QAT) & 16.47&  34.21&  16.43 & 35.69 \\
			\midrule
			5-bit (S8BQAT) & 8.14 & 21.44 & 8.64 & 22.35 \\
            \hspace{1pt}\text{w. } $\lambda$=$5e$-$6$, $\tau$=$35$ & 8.37 & 21.63 & 8.81 & 22.80 \\
            \hspace{1pt}\text{w. } $\lambda$=$5e$-$5$, $\tau$=$35$ & 8.20 & 21.52 & 8.78 & 22.45 \\
            \hspace{1pt}\text{w. } $\lambda$=$5e$-$2$, $\tau$=$35$ & 8.73 & 21.74 & 9.01 & 22.54 \\
            \hspace{1pt}\text{w. } $\lambda$=$5e$-$4$, $\tau$=$5$ & 8.16 & 21.46 & 8.71 & 22.39 \\	
            \hspace{1pt}\text{w. } $\lambda$=$5e$-$4$, $\tau$=$120$ & 9.10 & 21.84 & 9.34 & 22.37 \\
            4-bit (S8BQAT) & 8.79 & 21.79 & 8.94 & 22.65 \\
			\bottomrule
		\end{tabular}
	\label{tab:ablation}
	\vspace{-0.17in}
\end{table}

\begin{figure}[t]
	\centering
{\includegraphics[width=\linewidth]{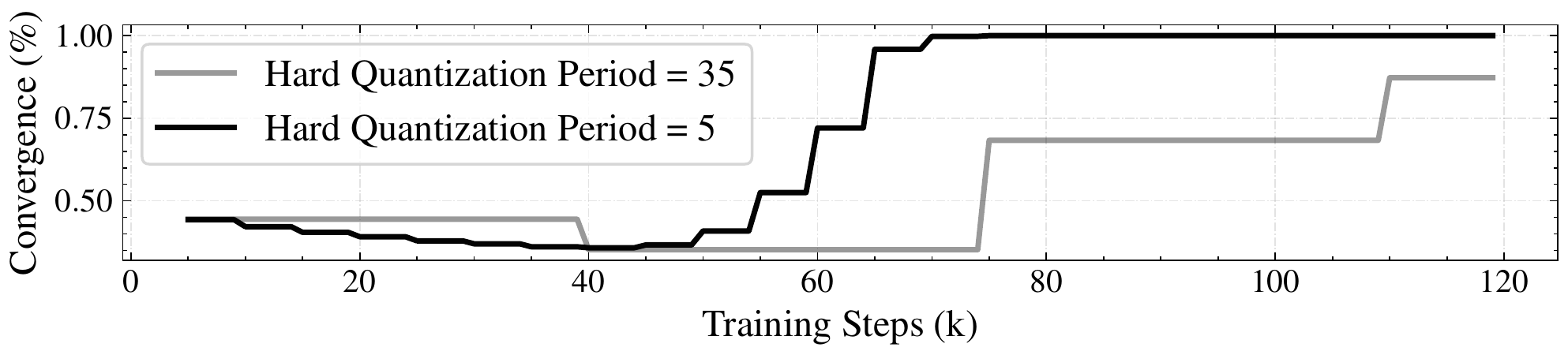}}
	\vspace{-0.15in}
	\caption{The average quantization convergence rate ($\gamma$) increases with periodic hard compressor.}
	\label{fig:conv}
	\vspace{-0.19in}
\end{figure}

\subsubsection{Ablation analysis}

We also conduct the ablation analysis (Table \ref{tab:ablation}) by alternating the \hieng{regularization} weight ($\lambda$) for the soft regularizer and the hard compression period ($\tau$). 
\hieng{The baseline models have M-I architecture and linear-QAT with 8, 5, and 4-bit weights, i.e. the model is QAT trained and post-training quantized with x bits in which x=$\{8,5,4\}$. For S8BQAT models, we consider 4-bit and 5-bit,  training/compressing and post-training quantizing all layers. All models} are trained on LibriSpeech dataset with all other training hyperparameters being the same.

As shown in Table \ref{tab:ablation}, the linear-QAT method \cite{nguyen2020quantization} performs well in 8-bit, but fails to preserve the accuracy level in sub-8-bit modes.  In contrast, the proposed soft compressor leverages Lloyd-Max scalar quantization theory, distilling near-optimal non-linearly spaced centroids from the \zhenk{32-bit} model. Consequently, it leads to no accuracy loss when the model is \hieng{quantized into 5-bit}. Even in 4-bit, our method achieves less than 3\% relative WER degradation in test sets and dev-other set, 
significantly smaller than linear 4-bit
QAT, i.e. the competing method. We use $\lambda$=$5e$-$4$ and $\tau$=$35$ for the 4-bit and 5-bit models with S8BQAT.
\hieng{Note that a large regularization weight ($\lambda$=$5e$-$2$) affects the model performance negatively, while smaller regularization weights (less than $5e$-$4$) lead to small degradation if the hard compressor is included.}

\hieng{It is observed that the hard compressor is essential} as the soft compressor 
\hieng{by itself needs a large number of training steps to get high convergence rates}: if the hard compressor is removed during training, WER degradation becomes noticeable especially for dev-clean and test-clean datasets \zhenk{(see the row with $\tau$=$120$ in Table \ref{tab:ablation})}. 
This phenomenon is also observed in Figure \ref{fig:conv}: in the beginning phase, the learning rate is relatively large, and even with periodic hard compressor, the convergence rate $\gamma$ is below 0.5 as model weights actively shift to lower the accuracy loss. As the weights are \hieng{gradually} converged, the hard compressor starts to effectively increase $\gamma$. Empirically, $\tau$=35 yields slightly better performance than $\tau$=5, indicating that a too frequent hard compressor may restrain weight update.

\section{Conclusion}
\label{sec:con}
\hieng{We proposed a novel sub-8-bit quantization aware training (S8BQAT) mechanism and confirmed its effectiveness via speech recognition task. It consists of a Lloyd-
Max quantizer-like soft compressor that distills near optimal quantization centroids from \zhenk{32-bit} weights, along with a periodic hard compressor which mimics post-training/runtime compression. With S8BQAT, we demonstrate that it is possible to reduce both the WER and on-device latency by 16\% and 5\%, respectively. The proposed method is applicable to various neural network based on-device speech processing tasks.}

\bibliographystyle{IEEEbib}
\bibliography{refs.bib}

\end{document}